\documentclass{article}
\usepackage{iclr2025_conference,times}

\usepackage{url}

\usepackage{times}
\usepackage{soul}
\usepackage{url}
\usepackage[hidelinks]{hyperref}
\usepackage[utf8]{inputenc}
\usepackage{caption}
\usepackage{graphicx}
\usepackage{booktabs}
\usepackage{amsmath}
\usepackage{amsthm}
\usepackage{booktabs}
\usepackage{algorithm}
\usepackage[printonlyused]{acronym}
\usepackage[switch]{lineno}
\usepackage{amsfonts}
\usepackage{amssymb}
\usepackage{cleveref}
\usepackage{xcolor}
\usepackage{setspace}
\usepackage{subcaption}
\usepackage{algorithmic}
\usepackage{multirow}
\usepackage{colortbl}
\theoremstyle{plain}

\theoremstyle{definition}

\theoremstyle{remark}

\definecolor{highlightgreen}{HTML}{009901}
\definecolor{highlightred}{HTML}{FD6864}
\definecolor{ao(english)}{rgb}{0.0, 0.5, 0.0}
\definecolor{arylideyellow}{rgb}{0.91, 0.84, 0.42}
\definecolor{LightCyan1}{rgb}{0.88,1,1}

\newcommand{\bb}[1]{\textbf{#1}}
\usepackage[textsize=tiny]{todonotes}
\acrodef{tta}[TTA]{text-to-audio}
\acrodef{vta}[VTA]{video-to-audio}
\acrodef{method}[PPAE]{Prompt-guided Precise Audio Editing}
\acrodef{ptp}[PTP]{Prompt-to-Prompt}
\acrodef{llm}[LLM]{Large Language Models}
\acrodef{clap}[CLAP]{Contrastive language-audio pretraining}
\acrodef{gan}[GAN]{Generative Adversarial Network}
\acrodef{ldm}[LDM]{Latent Diffusion Model}
\acrodef{vae}[VAE]{variational auto-encoder}
\acrodef{ddim}[DDIM]{Denoising Diffusion Implicit Models}
\acrodef{ddpm}[DDPM]{Denoising Diffusion Probabilistic Models}
\acrodef{fd}[FD]{Fréchet distance}
\acrodef{fad}[FAD]{Fréchet audio distance}
\acrodef{sd}[SD]{Spectral distance}
\acrodef{kl}[KL]{Kullback–Leibler}
\acrodef{is}[IS]{Inception Score}
\acrodef{clip}[CLIP]{Contrastive Language-Image Pre-training}
\acrodef{tts}[TTS]{test-to-sound}
\acrodef{vaa}[VAA]{Video-Audio Alignment}
\acrodef{stft}[STFT]{Short-Time Fourier Transform}
\acrodef{pam}[PAM]{Prompting Audio-Language Models}

\newcolumntype{x}{>{\columncolor{LightCyan1}}c}

\title{Video-to-Audio Generation \\
with Hidden Alignment}

\usepackage{hyperref}
\hypersetup{
    colorlinks = false,
    linkbordercolor = {0 1 0}, 
    pdfborderstyle={/S/S/W 1} 
}

\author{
    Manjie Xu$^{1,2}$, Chenxing Li$^{1, *}$, Xinyi Tu$^{3}$, Yong Ren$^{1,4}$,  Rilin Chen$^{1}$, Yu Gu$^{1}$, Wei Liang$^{2, *}$, Dong Yu$^{5, *}$ \\
    \small $*$ corresponding authors\\
    \small $^1$ Tencent AI Lab, Beijing, China\\
    \small $^2$ Beijing Institute of Technology\\
    \small $^3$ University of California, Berkeley\\
    \small $^4$ Institute of Automation, Chinese Academy of Sciences, Beijing, China\\
    \small $^5$ Tencent AI Lab, Seattle, USA\\
}

\iclrfinalcopy 

\begin{document}
\maketitle
\begin{abstract}

Generating semantically and temporally aligned audio content in accordance with video input has become a focal point for researchers, particularly following the remarkable breakthrough in text-to-video generation. In this work, we aim to offer insights into the video-to-audio generation paradigm, focusing on three crucial aspects: vision encoders, auxiliary embeddings, and data augmentation techniques. Beginning with a foundational model built on a simple yet surprisingly effective intuition, we explore various vision encoders and auxiliary embeddings through ablation studies. Employing a comprehensive evaluation pipeline that emphasizes generation quality and video-audio synchronization alignment, we demonstrate that our model exhibits state-of-the-art video-to-audio generation capabilities. Furthermore, we provide critical insights into the impact of different data augmentation methods on enhancing the generation framework's overall capacity. We showcase possibilities to advance the challenge of generating synchronized audio from semantic and temporal perspectives. We hope these insights will serve as a stepping stone toward developing more realistic and accurate audio-visual generation models.
\end{abstract}

\section{Introduction}
We humans perceive visual and audio input as two complementary aspects of our surroundings. When we witness an event, we instinctively expect to hear the corresponding sounds as we see it unfold. This integrated perception of sight and sound enhances our understanding and interpretation of the world around us. Modern video generation models, however, typically focus on generating visual content based on a set of textual prompts \citep{blattmann2023stable,zeng2023make,bartal2024lumiere}. These models often lack the ability to incorporate audio information, resulting in outputs that appear as inferior imitations rather than realistic representations.

Our objective is to generate semantically and temporally-aligned audio content for a given silent video. This ambitious goal is rooted in extensive research conducted in the \ac{tta} domain, where researchers have proposed several robust baselines for audio generation~\citep{kreuk2022audiogen,liu2023audioldm,ghosal2023text}. Furthermore, there have been recent attempts at \ac{vta} tasks~\citep{sheffer2023hear,luo2024diff,v2a-mapper}. While these works have made significant strides, they still fall short of natural and arbitrary audio generation.

A common trait amongst these baseline models is their utilization of non-autoregressive generation frameworks, particularly \ac{ldm}~\citep{rombach2022high}. These models generate audio content by leveraging textual or visual features as generative conditions. However, the distinction between \ac{vta} and \ac{tta} presents two key challenges: 1) ensuring semantic coherence with the input condition, and 2) ensuring temporal alignment between the generated audio and the video. This is due to the necessity of accurately extracting and interpreting visual features that produce sounds, as well as identifying when these features generate sounds, even if they are continuously present in the scene.

\begin{figure*}[t!]
    \centering
    \small
    \includegraphics[width=\linewidth]{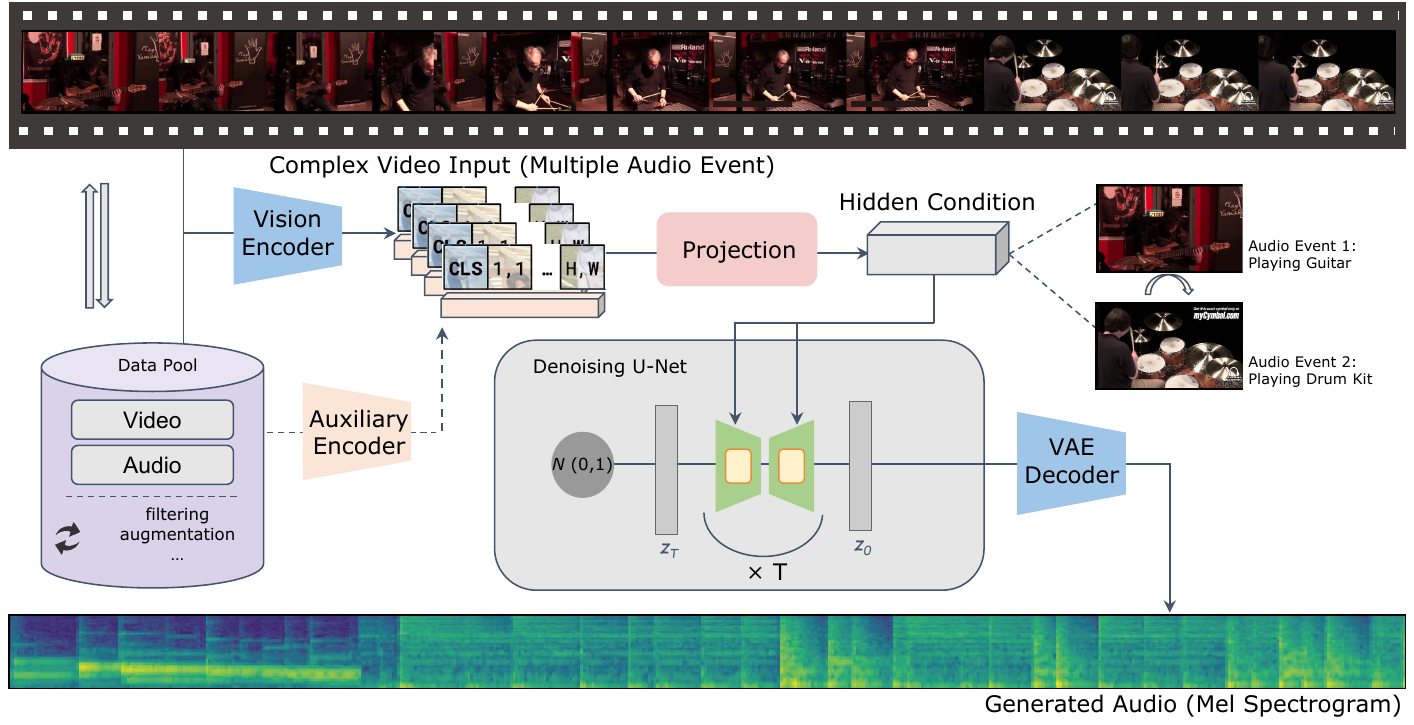}
    \caption{\textbf{Overview of the VTA-LDM framework.}  Given the silent video, our model generates semantically-related and temporally-aligned audios that accurately correspond to the visual events. The framework is based on a \ac{ldm} with encoded vision features as the generation condition.}
    \label{fig:tta_intro}
\end{figure*}

We try to provide insights into the \ac{vta} training paradigm by focusing on three key aspects: 1) the vision encoder, 2) the auxiliary embedding, and 3) the data augmentation. Vision encoder is responsible for extracting and interpreting visual features from the input video, capturing complex visual patterns that are integral to generating relevant audio content. It also has the potential to encode temporal information. Auxiliary embedding serves as a valuable source of additional contextual information for the model, e.g. textual description, position embedding, or other metadata associated with the video. Data augmentation can help improve the model's generalization capabilities by introducing variations and perturbations in the training data. In particular, time-stretching of the training data can create fast-switching video scenes, which can push the model to better learn and understand temporal relations between visual and auditory information.

Our study initiates from training a foundational \ac{vta} model, denoted as VTA-LDM, drawing upon the successful text-to-audio \ac{ldm} framework. By employing a Clip-based vision encoder~\citep{radford2021learning}, we concatenate frame-level video features temporally and map them using a projector as the generation condition. The overall framework of VTA-LDM is shown in \cref{fig:tta_intro}. The evaluation of these models concentrates on two primary aspects: semantic alignment and temporal alignment. Semantic alignment assesses the generated output's semantic coherence and audio relevance, while temporal alignment gauges the synchronization between the generated audio and corresponding video frames. We then conduct comprehensive ablation experiments to examine the three key factors that can influence the overall performance. Results show that even our basic framework achieves state-of-the-art results in \ac{tta} tasks, generating semantically and partly-temporally aligned audio content based on visual input. Furthermore, the integration of additional features into the framework can significantly enhance the generation quality and the synchronization between audio and visual elements from various perspectives. For instance, the inclusion of extra textual prompts can aid in semantic understanding, positional embedding and concatenation augmentation can improve temporal alignment, and pretraining can enhance the overall quality of generation. These findings indicate that the overall framework empowers the model to learn and comprehend the dynamics of the scene and demonstrates the potential for further exploration and refinement.

\begin{enumerate} 
    \item We introduce an effective \ac{vta} framework, named VTA-LDM, which attains state-of-the-art performance in \ac{vta} tasks. 
    \item We investigate three fundamental aspects within the \ac{vta} paradigm, providing valuable insights into model design and training processes: 1) a linear combination of semantic video features is sufficient for encoding information required for aligned audio generation; 2) additional auxiliary embeddings assist in filtering out chaotic visual information and focusing on crucial elements; 3) pre-training enhances overall generation capability, and filtering high-quality data improves audio-video alignment.
    \item Through numerous ablation experiments and evaluations, we supply detailed quantitative results for \ac{vta} tasks.
\end{enumerate}

We hope our experiments and results lay a solid foundation for future research in \ac{vta} field.

\section{Related Work}
\subsection{Diffusion-based Generation}
The success of audio generation tasks primarily stems from \ac{tta} diffusion models. These diffusion models were initially employed in image generation tasks~\citep{ho2020denoising, dhariwal2021diffusion, nichol2021glide, rombach2022high}, and then extended to audio and video generation~\citep{liu2023audioldm, liu2023audioldm2, singer2022make, 
blattmann2023stable,videoworldsimulators2024}. Compared with previous works like Audiogen~\citep{kreuk2022audiogen}, which are predominantly based on regressive generative models, diffusion-based models excel in generation quality and diversity. For instance, Audioldm~\citep{liu2023audioldm} learns continuous audio representations in the latent space. Tango~\citep{ghosal2023text} incorporates an instruction-tuned Large Language Model (LLM) FLAN-T5 to leverage its powerful representational capabilities. Other studies~\citep{liu2023audioldm2, huang2023make2, luo2024diff, zeng2023make, xu2024prompt} also utilize diffusion models, in conjunction with other enhancing methods, to improve the generation capacity.

\subsection{Multimodal Audio Generation}
Numerous studies have concentrated on multi-modal generation tasks related to the audio modality, extending beyond text and conducting significant experiments across various modalities. Previous research~\citep{iashin2021taming}, has focused on multi-class visually guided sound synthesis utilizing VQGAN~\citep{esser2021taming}. Recent work includes IM2WAV~\citep{sheffer2023hear}, which employs a two-stage audio generation pipeline to generate low-level audio representation autoregressively, followed by high-level audio tokens from images. Similarly, Diff-Foley~\citep{luo2024diff} uses contrastive audio-visual pretraining to learn temporally aligned features for synthesizing audio from silent videos, drawing on \ac{ldm}. Seeing\&Hearing~\citep{xing2024seeing} and T2AV~\citep{mo2024text} show open-domain visual-audio generation capability based on similar multi-modality latent aligners. Other studies~\citep{kurmi2021collaborative, ruan2023mm, yariv2024diverse} have attempted joint video-audio generation using the same or aligned latent. Despite these advancements, generating synchronized audio that corresponds with a given video remains a significant challenge, particularly at the audio event level.

\section{Baseline Framework}

\subsection{Overall Framework}
Drawing inspiration from previous \ac{tta} works~\citep{liu2023audioldm, ghosal2023text} and multi-modal research~\citep{luo2021clip4clip, sheffer2023hear}, we develop a fundamental \ac{vta} framework VTA-LDM consisting of several key components: a vision encoder, a conditional \ac{ldm}, and a mel-spectrogram/audio \ac{vae}. Specifically, we utilize vision features extracted from pre-trained vision encoders and feed them to the \ac{ldm} as the generation condition through a linear projection layer. The \ac{ldm} operates on the latent audio representation of the mel-spectrogram. The pre-trained audio \ac{vae} assists in decoding the denoised latent output to a mel-spectrogram, which is then fed to a vocoder to generate the final audio.

\subsection{Vision Encoder}
Vision Encoder is responsible for encoding not only the semantic meanings of the video $V$ but also the temporal information required for alignment with the generated audio. We employ pretrained vision encoders $f_V$ such as CLIP4CLIP~\citep{luo2021clip4clip} to extract the visual features from the input video. These features capture the essential visual information, including objects, actions, and scene context. We use a projection layer $\phi$ to map these features to the desired dimension of the diffusion condition. The vision encoders are all frozen during the training process, while the projection layer will be trained from scratch.

\subsection{\acf{ldm}}
Given the original input $x_0$, diffusion models follow a Markov chain of diffusion steps to gradually add random noise to data until it follows a Gaussian distribution ${N}(0, I)$, represented by $x_t$. The model then learns the reverse de-noising process to recover the original input data. For computational efficiency, \ac{ldm} incorporates a trained perceptual compression model to encode the input $x$ into a low-dimensional latent space $Z$. In text-based generation tasks, the generation condition is often a given text description. In our \ac{vta} implementation, the condition $c$ is the projected video feature $\phi(f_V)$. To perform sequential de-noising, we train a network$\epsilon_\theta$ to predict artificial noise, following the objective:
\begin{equation}
\min_\theta \mathbb{E}_{z_0, \epsilon \sim \mathcal{N}(0, I), t \sim \text{Uniform}(1, T)} \lVert \epsilon - \epsilon_\theta(z_t, t, \phi(f_V)) \rVert_2^2,
\end{equation}
where the condition $C = \phi(f_V)$ is the projected visual embedding of the input video. We employ a classifier-free generation approach in the reverse process, which has been shown to be an effective method for text-guided generation and editing~\cite{ho2022classifier}. Given a latent variable and a textual prompt, the generation is performed both conditionally and unconditionally, and then weighted according to a given scale. Formally, let $\varnothing$ be the null text embedding, $w$ denotes the guidance scale, and the generation process can be defined by:
\begin{equation}
\tilde{\epsilon}_\theta = w \cdot \epsilon_\theta(z_t, t, \phi(f_V)) + (1 - w) \cdot \epsilon_\theta(z_t, t, \varnothing).
\end{equation}
In text-based generation, $\varnothing$ is commonly defined as $\phi(\text{" "})$ to represent the null condition. In \ac{tta} tasks, we use a zero embedding to serve as the $\phi(NULL)$, representing the null visual condition.
\subsection{Audio VAE and Vocoder}
The audio source is firstly conveyed to the mel-spectrogram using \ac{stft}, and then to the latent representation $z$ with a pre-trained audio \ac{vae}. We use the pre-trained \ac{vae} from AudioLDM \citep{liu2023audioldm}. In the reverse process, the audio is recovered from the latent $z$ using the Hifi-GAN vocoder \citep{kong2020hifi}.

\section{Experimental Setup}
\subsection{Dataset}
Our primary experiments are conducted on VGGSound~\cite{chen2020vggsound}, a dataset comprising over 550 hours of videos with acoustic visual-audio event pairs. We train our models on 200k videos and evaluate them on 3k videos. For certain ablations, we either construct an augmented test set based on the original dataset by randomly cutting and combining segments, or import large unlabelled audio or video corpora for pre-training purposes. More details regarding the dataset preparation, training, and evaluation procedures can be found in \cref{sec:seven} and \cref{sec:eight}. 

\subsection{Objective Evaluation}
We utilize commonly used quantitative metrics for the objective evaluation. For evaluating semantic consistency, \ac{fd}, \ac{fad}, \ac{is}, \ac{kl} measure the semantic similarity of the generated audios with the ground-truth target in distribution, as well as the generation diversity and quality of the audios. \ac{clap}~\citep{laionclap2023} score calculates how well the generated audio aligns with the textual description of the audio. For evaluating temporal alignment, AV-Align~\citep{yariv2024diverse} assesses the alignment of generated audio with the input video. CAVP~\citep{luo2024diff} evaluates how well the audio aligns with the visual input at both semantic and temporal levels. Besides, \ac{pam}~\citep{deshmukh2024pam} assesses the quality of generated audio.

\subsection{Subjective Evaluation}
We include a subjective evaluation of our models, assessing overall quality, audio quality, video-audio semantic alignment, and video-audio temporal alignment using a scale of 1 to 100. In the subjective evaluation, we combine the generated audio with the original video pieces, and show the combined video to the human participants without telling them the concrete test cases. Detailed definition of these metrics can be found in \cref{sup_subjective}. 

\section{Experiments}
\subsection{Basic Configs}
Our \ac{ldm} structure is configured with a cross-attention dimension of 2048 and features 8 input and output channels. We train our models with 300 warmup steps, a learning rate of 6e-5, and a batch size of 128. All models are trained for a total of 120 epochs. All the models are trained on 8 Nvidia A100 cards. During the inference steps, we set the denoise steps to 300, and the number of samples per audio 1. We utilize classifier-free guidance with a guidance scale of 3.
\subsection{Initial Results}
We first compare our foundational model with other existing \ac{vta} baselines to show its capability in audio generation, including  IM2WAV~\citep{sheffer2023hear}, Diff-Foley~\citep{luo2024diff}, FoleyCrafter \citep{zhang2024foleycrafter}, Seeing\&Hearing~\citep{xing2024seeing} and T2AV~\citep{mo2024text}. These compared baselines primarily utilize the diffusion framework. Although there are additional \ac{vta} baselines available, we consider these to be representative and indicative of SOTA performance. We also list the vision encoders these baselines use for reference. For certain benchmarks, we omit their alignment metrics, as they are tailored to produce shorter audio segments, which does not align with competitive standards. See \cref{supp_baselines} for more details about these baselines.

\begin{table}[!ht]
    \small
    \caption{\textbf{Comparison with other baselines.}}
    \rowcolors{2}{LightCyan1}{}
    \centering
    \resizebox{0.9\linewidth}{!}{
    \begin{tabular}{cccccccccc}\toprule
                   & \textbf{VE} & \textbf{FAD} {\color{highlightgreen} $\downarrow$}  & \textbf{IS} {\color{highlightred} $\uparrow$} & \textbf{FD} {\color{highlightgreen} $\downarrow$} & \textbf{KL} {\color{highlightgreen} $\downarrow$} & \textbf{PAM} {\color{highlightred} $\uparrow$} & \textbf{CLAP} {\color{highlightred} $\uparrow$} & \textbf{CAVP} {\color{highlightred} $\uparrow$} & \textbf{AV-Align} {\color{highlightred} $\uparrow$}  \\\midrule
    GT & - & - & - & - & - & \textit{\bb{0.316}} & \textit{\bb{0.492}} & \textit{\bb{0.813}} & \textit{\bb{0.237}} \\ 
    IM2WAV & CLIP & 6.32 & 7.46 & 42.13 & 2.38 & 0.182 & 0.306 & 0.781 & 0.192 \\
    Diff-Foley & CAVP & 7.32 & 9.62 & 41.09 & 6.03 & 0.226 & 0.441 & \textbf{0.802} & 0.187 \\
    FoleyCrafter & CLIP & 4.44 & 9.44 & 27.00 & 4.57 & \textbf{0.307} & 0.179 & 0.762 & 0.239 \\
    Seeing\&Hearing& ImageBind & 7.32 & 5.83 & 32.92 & 2.62 & - & - & - & - \\
    T2AV&  VA-CLAP & 4.05 & 8.02 & 33.29 & \bb{2.12} & - & - & - & - \\
    VTA-LDM (base) & CLIP4Clip & 2.05 & 10.10 & 25.50 & 3.81 & 0.245 & 0.452 & 0.800 & 0.225\\
    VTA-LDM+Text+Concat & CLIP4Clip & \bb{2.12} & \bb{11.21} & \bb{21.10} & 2.42 & 0.285 & \textbf{0.484} & 0.800 & \bb{0.242}\\
    \bottomrule
    
    \end{tabular}
    }
    \label{tab:baseline}
\end{table}

Table~\ref{tab:baseline} presents the results of our foundational model compared to existing \ac{vta} baselines with different vision encoders (VE), demonstrating its capability in audio generation\footnote{For Seeing\&Hearing and T2AV, as the codes are not publicly released, we adopt the metrics from their original papers.}. We observe that our model is competitive with all of the baselines across the metrics, showcasing its superiority in generating high-quality, diverse, and temporally-aligned audios. Specifically, our model achieves a lower \ac{fad} and \ac{fd} compared to the IM2WAV and Diff-Foley baseline, indicating better distribution similarity between generated and ground truth audios. Regarding CLAP, CAVP, and AV-Align, our model exhibits superior performance compared to the baselines, emphasizing its ability to generate semantically-related and temporally-aligned audios. 

\subsection{Experiments on Vision Encoders}

\begin{table}[!ht]
\small
\caption{\textbf{Ablation on vision encoders.} }
\centering
\rowcolors{2}{LightCyan1}{}
\resizebox{\linewidth}{!}{

\begin{tabular}{cccccccccc}\toprule
    & Training Data & \textbf{FAD} {\color{highlightgreen} $\downarrow$}  & \textbf{IS} {\color{highlightred} $\uparrow$} & \textbf{FD} {\color{highlightgreen} $\downarrow$} & \textbf{KL} {\color{highlightgreen} $\downarrow$} & \textbf{PAM} {\color{highlightred} $\uparrow$} & \textbf{CLAP} {\color{highlightred} $\uparrow$} & \textbf{CAVP} {\color{highlightred} $\uparrow$} & \textbf{AV-Align} {\color{highlightred} $\uparrow$}  \\\midrule
CLIP4Clip & Howto100M & \bb{2.05} & \bb{10.10} & \bb{25.50} & \bb{3.81} & 0.245 & \bb{0.452} & \bb{0.800} & \bb{0.225}\\
ImageBind  & AudioSet(2M) & 8.40 & 6.85 & 42.18 & 8.41 & 0.166 & 0.186 & 0.791 & 0.189 \\
LanguageBind  & VIDAL-10M & 2.94 & 9.03 & 28.65 & 4.76 & 0.201 & 0.338 & 0.798 & 0.200\\
ViViT & Kinetics 400 & 15.99 & 3.88 & 58.14 & 7.33 & \bb{0.288} & 0.223 & 0.766 & 0.189 \\
V-JPEA & VideoMix2M & 14.28 & 4.27 & 60.04 & 11.89 & 0.142 & 0.080 & 0.786 & 0.186\\
CAVP & AudioSet(2M)+VGGSound(200k) &  4.27 & 4.83 & 56.77 & 3.88 & 0.192 & 0.137 & 0.787 & 0.220 \\\bottomrule

\end{tabular}}
\label{tab:vis_enc}
\end{table}

We begin by investigating how different vision encoders may influence the \ac{vta} results. Our study includes popular vision encoders such as Clip4Clip~\citep{luo2021clip4clip}, Imagebind~\citep{girdhar2023imagebind}, LanguageBind~\citep{zhu2023languagebind}, V-JEPA~\citep{bardes2024vjepa}, ViViT~\citep{arnab2021vivit}, and CAVP~\citep{luo2024diff}. These vision encoders can primarily be categorized into two groups: 1) image encoders trained with multimodal alignment to enhance semantic understanding (\textit{e.g.}, Clip4Clip and Imagebind), and 2) video feature encoders designed to extract semantic and temporal information from videos. Details can be found in \cref{supp_encoder}.

We show the results in \cref{tab:vis_enc}. The experiments highlight the influence of different vision encoders on the video-to-audio generation task. The choice of vision encoder plays a critical role in capturing the semantic and temporal information from the input videos, which directly impacts the quality, diversity, and alignment of the generated audio. From the results, it is evident that Clip4Clip, which is based on the pre-trained model Clip, outperforms other methods in generating high-quality and diverse semantically-related audio content. On the other hand, methods like ViViT and V-JPEA, which utilize different vision encoders, show comparatively lower performance in most metrics. While in these cases, CAVP achieves better temporal alignment as shown by the AV-Align score and the CAVP score.

\begin{figure*}[h!]
    \centering
    \small
    \includegraphics[width=\linewidth]{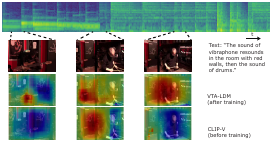}
    \caption{\textbf{The saliency map of our model's interest upon the visual input.} We illustrate that VTA-LDM has the ability to learn and concentrate on potential objects capable of producing sound. Furthermore, the model is designed to focus on various sections of the frame across different time intervals, although the attention is calculated based on the final audio latent representation only once. }
    \label{fig:sailency}
\end{figure*}

To further demonstrate how the model's generation part engagements with the input video, we compute the saliency map of the input visual feature based on the generated audio content, aiming to empirically show how VTA-LDM processes the input video frames. As illustrated in \cref{fig:sailency}, our model focuses on different potential video regions that could potentially be the source of the audio. The model can engage in an attention shuffle with the frames as they progress, indicating its ability to capture temporal information between the frames, despite using the pure semantic vision encoder CLIP. For a safe conclusion, we note that a simple temporal combination of semantic vision features of the video frames may be sufficient to capture the features from the video in \ac{vta} tasks, eliminating the need for re-training a video feature encoder.

\subsection{Experiments on Auxiliary Embeddings}
\label{sec:seven}

Auxiliary embeddings serve as additional information beyond visual features, potentially enhancing video understanding and anchoring audio events to improve the generation process. We explore the integration of extra information from various modalities, including textual descriptions that encapsulate the overall video content, position embeddings that indicate the sequence of frames, and optical flow that captures pixel-level changes between frames within the video. Details can be found in \cref{supp_aux}.

\begin{table}[!ht]
\small
\caption{\textbf{Ablation on auxiliary embeddings.} }
\centering
\rowcolors{2}{LightCyan1}{}
\resizebox{\linewidth}{!}{
\begin{tabular}{ccccccccc}\toprule
               & \textbf{FAD} {\color{highlightgreen} $\downarrow$}  & \textbf{IS} {\color{highlightred} $\uparrow$} & \textbf{FD} {\color{highlightgreen} $\downarrow$} & \textbf{KL} {\color{highlightgreen} $\downarrow$} & \textbf{PAM} {\color{highlightred} $\uparrow$} & \textbf{CLAP} {\color{highlightred} $\uparrow$} & \textbf{CAVP} {\color{highlightred} $\uparrow$} & \textbf{AV-Align} {\color{highlightred} $\uparrow$}  \\\midrule
Text-only & \bb{1.98} & 10.02 & 21.25 & \bb{2.39} & 0.278 & \bb{0.486} & 0.764 & 0.176 \\
Vision-only(VTA-LDM) & 2.05 & 10.10 & 25.50 & 3.81 & 0.245 & 0.452 & 0.800 & 0.225\\
Vision+Text & 2.09 & 10.61 & \bb{21.13} & 2.47 & \bb{0.284} & \bb{0.486} & \bb{0.802} & \bb{0.228}
\\
Vision+PE & 2.36 & 8.16 & 29.71 & 4.88 & 0.250
 & 0.332 & 0.801 & 0.223
 \\
Vision+Text+PE & 2.99 & \bb{12.71} & 25.98 & 2.76 & 0.280 & 0.476 & 0.794	 & 	0.189 \\
Vision+Optical Flow & 3.02 & 9.88 & 28.32 & 3.01 & 0.230 & 0.473 & 0.792 & 0.224 \\\bottomrule

\end{tabular}}
\label{tab:add_emb}
\end{table}

Results in \cref{tab:add_emb} show that auxiliary embeddings, such as extra text embeddings and extra position embeddings, provide additional information that can enhance the model's understanding of the input data and improve the quality of generated audios. Adding extra text embeddings improves the Inception Score (IS) from 10.10 to 10.61 and reduces the \ac{fad}, \ac{fd}, and \ac{kl}, indicating that the additional semantic information from the text can enhance the quality and diversity of the generated audios, while adding extra position embeddings (PE) appears to have a mixed effect. Combining both text and position embeddings yields the highest IS score of 12.71 but also increases the FAD and reduces the AV-Align score. This suggests that while the combined embeddings can enhance the diversity of the generated audios, they might not necessarily improve the alignment with the ground truth audios.

\begin{figure*}[h!]
    \centering
    \small
    \includegraphics[width=\linewidth]{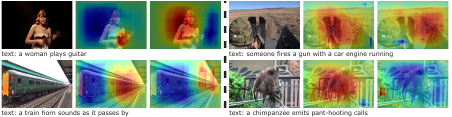}
    \caption{\textbf{A Comparison Between Models Without Additional Text Embedding.} The left saliency maps encode the text embedding, while the right ones do not. We demonstrate that extra text embeddings can aid the model in gaining a deeper understanding of the visual content.}
    \label{fig:app_lan}
\end{figure*}

We observe that additional embeddings, particularly extra text embeddings, can not only assist the model in comprehending the input video but also concentrate on the actual object producing the audio, thereby improving audio-visual alignment, as illustrated in \cref{fig:app_lan}. Visual inputs are more complex and chaotic, making it challenging to discern the true source of the audio. For instance, (on the top left) a woman speaks without moving her mouth, but a non-fine-grained vision encoder cannot recognize this. The extra textual input can help identify the real focus, which is the "guitar."

\subsection{Experiments on Data Augmentation}
\label{sec:eight}

\begin{table}[!ht]
\small
\caption{\textbf{Ablation on data augmentation.} }
\centering
\rowcolors{2}{LightCyan1}{}
\begin{tabular}{ccccccccc}\toprule
               & \textbf{FAD} {\color{highlightgreen} $\downarrow$}  & \textbf{IS} {\color{highlightred} $\uparrow$} & \textbf{FD} {\color{highlightgreen} $\downarrow$} & \textbf{KL} {\color{highlightgreen} $\downarrow$} & \textbf{PAM} {\color{highlightred} $\uparrow$} & \textbf{CLAP} {\color{highlightred} $\uparrow$} & \textbf{CAVP} {\color{highlightred} $\uparrow$} & \textbf{AV-Align} {\color{highlightred} $\uparrow$}  \\\midrule
None(VTA-LDM) & \bb{2.05} & 10.10 & 25.50 & 3.81 & 0.245 & 0.452 & 0.800 & 0.225\\
Data Alignment Filter & 3.85 & 8.19 & 33.43 & \bb{0.95} & \bb{0.314} & \bb{0.484} & \bb{0.802} & \bb{0.281} \\
Concat Augment& 2.47 & 10.82 & 25.09 & 3.59 & 0.236 & 0.393 & 0.798 & 0.248 \\
Pretrain-Video & 2.93 & 9.90 & 32.10 & 1.28 & 0.292 & 0.267 & 0.769 & 0.262 \\
Pretrain-Audio & 2.48 & \bb{12.23} & \bb{23.61} & 1.38 & 0.232 & 0.469 & 0.798 &  0.250 \\\bottomrule

\end{tabular}
\label{tab:data_aug}
\end{table}

The data used in training is also crucial. We explore several approaches from different perspectives. Initially, we filter the training set using video-audio alignment labels to obtain high-quality video-audio pairs. We use a CLAP~\citep{laionclap2023} model to help select audio-video pairs with similar semantics based on extra textual labels (with $score>0.3$). We also use the AV-Align score to filter unmatched video-audio pairs (with $score>0.2$). Furthermore, we utilize data augmentation techniques to enhance the diversity and complexity of the training set. We randomly combine video and audio segments to construct a diverse and multi-content set of training data points. Additionally, we leverage extra training sets for pretraining and perform fine-tuning on the foundational model. These approaches are separately studied to illustrate their influence on the final generation results. More details can be found in \cref{supp_dataaug}.

The influence of different data augmentation methods on the performance of the \ac{vta} generation model can be observed in \cref{tab:data_aug}. We compare the base model without augmentation to models utilizing various data augmentation techniques. Data Clean leads to the most significant performance improvement. Removing noisy or irrelevant video-audio samples results in improvements mainly in audio quality and video-audio alignment. The drop in generation quality can be attributed to the decrease in the number of training instances. Concat Augment leads to an improvement in IS, suggesting that this technique can enhance the diversity of the generated audio content, as well as a significant improvement in AV-Align, indicating better video-audio alignment. However, the distance metrics, such as FAD and KL, show a slight decrease with these data augmentation methods. It is worth noting that while data augmentations within the data corpus can improve diversity and alignment, there may be a trade-off in terms of audio content itself. Pretraining with extra data is a natural idea that can help address this issue. The results show that pretraining on extra unlabelled video and audio data can help improve some generation metrics. We believe an appropriate combination of these augmentation methods can enhance the overall performance of the \ac{vta} task.

\section{Further Experiments \& Discussion}

\subsection{Subjective Evaluation}

\begin{table}[!ht]
\small
\caption{\textbf{Subjective evaluation results.}}
\centering
\resizebox{\linewidth}{!}{
\rowcolors{2}{LightCyan1}{}
\begin{tabular}{ccccc}\toprule
                & \textbf{Audio Quality} {\color{highlightred} $\uparrow$} & \textbf{Semantic Alignment} {\color{highlightred} $\uparrow$} & \textbf{Temporal Alignment} {\color{highlightred} $\uparrow$} & \textbf{Overall Quality} {\color{highlightred} $\uparrow$} \\\midrule
GT & 91.06$\pm$8.87 & 91.14$\pm$12.54 & 89.66$\pm$14.04 & 91.06$\pm$10.62 \\
Diff-Foley & 48.45$\pm$11.44 &	47.03$\pm$28.45  &	57.17$\pm$28.20 &	46.26$\pm$19.72 \\
IM2WAV  & 75.01$\pm$16.77 & 	58.35$\pm$28.94  & 48.33$\pm$20.85  & 	54.45$\pm$21.59  \\
VTA-LDM & 72.52$\pm$20.71 & 	64.92$\pm$24.51 & 	43.50$\pm$25.52 & 	49.04$\pm$22.23
 \\
VTA-LDM+Text & 83.30$\pm$11.45 & 69.45$\pm$22.99 & 59.97$\pm$26.31  & 66.12$\pm$21.57  \\
VTA-LDM+Text+Concat & \bb{86.46}$\pm$11.48 & \bb{83.59}$\pm$16.69 &\bb{79.32}$\pm$17.64 &\bb{82.06}$\pm$15.52 \\\bottomrule

\end{tabular}}
\label{tab:data_sub}
\end{table}

\begin{figure*}[t!]
    \centering
    \small
    \includegraphics[width=\linewidth]{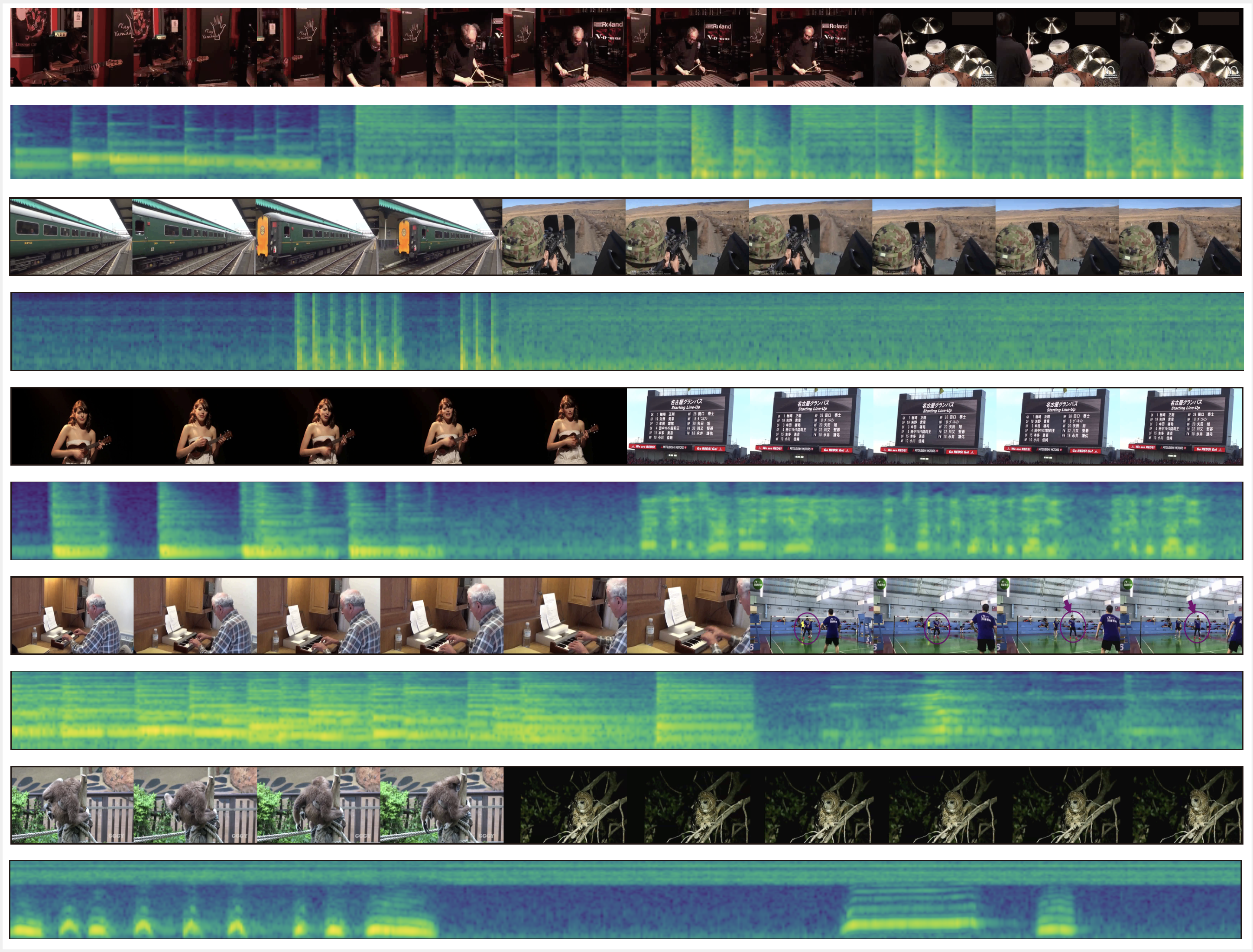}
    \caption{\textbf{Demos of the \ac{vta} generation.}  Given the silent video, our model generates semantically-related and temporally-aligned audios that accurately correspond to the visual events. }
    \label{fig:tta_demo}
\end{figure*}

\begin{figure*}[t!]
    \centering
    \small
    \includegraphics[width=\linewidth]{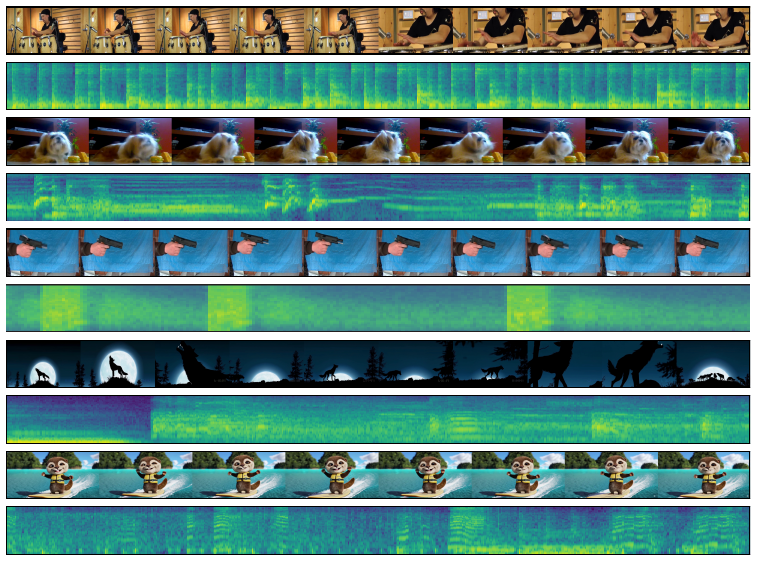}
    \caption{\textbf{Demos of the \ac{vta} generation on open-domain videos.} Videos are collected from YouTube or generated by OpenAI Sora~\citep{videoworldsimulators2024}. Although some test data points may exhibit styles distinct from those in the training dataset, our model concentrates on the semantic comprehension of video content and possesses a certain degree of out-of-domain generalization capability.}
    \label{fig:tta_ood}
\end{figure*}

We selected several ablation models with relatively better performance for the subjective evaluation to compare against the ground truth and other baselines. In each group, we randomly selected 20 video-audio pairs. The results are presented in Table~\ref{tab:data_sub}. 

In general, we observe that the subjective evaluations from our human participants align with the objective metrics, where our framework outperforms the other baselines. However, we also note that the performance still has room for improvement to reach the level of natural and realistic audio. Upon closer examination of the individual metrics, we observe that the VTA-LDM+Text+Concat model yields the best results among the ablation models, and also surpasses other baselines, indicating the effectiveness of combining auxiliary embeddings and data augmentation techniques. The improvements in semantic and temporal alignment, in particular, highlight the importance of incorporating these elements in the video-to-audio training paradigm. 

We also note that almost all baselines exhibit a significant variance. Upon further examination of the generated cases, we find that the models are sensitive to certain factors, such as the complexity of the video scene, the presence of multiple audio sources. From any level, there is still a noticeable gap between the performance of existing \ac{vta} models and the ground truth. 

\subsection{Overall discussion}
We show a branch of \ac{tta} demos in \cref{fig:tta_demo}. Based on the quantitative results above, we can safely conclude that our vanilla structure has demonstrated its capability to effectively address the semantic alignment of TTA tasks. The vanilla structure's design enables it to capture and translate the semantic information from the text prompts into the generated audio content, resulting in high-quality, semantically-aligned audio outputs. 

However, while the vanilla structure exhibits the potential to address the temporal alignment of TTA tasks, it only partially solves this problem. Temporal alignment, which involves aligning the generated audio events with the corresponding visual events in the video, is a more complex issue that requires additional considerations. We introduced additional modifications, such as position encoding and data augmentation, to enhance the overall performance. Position encoding helps the model better understand the temporal order of events, improving the temporal alignment of the generated audio content. Data augmentation, on the other hand, enhances the model's generalizability by exposing it to a wider variety of complex data scenarios.
While these modifications have generally led to improvements in both semantic and temporal alignment, they have also occasionally resulted in negative feedback, as shown in \cref{tab:add_emb} and \cref{tab:data_aug}.

\Cref{fig:tta_demo} shows that our model generally performs well in temporally generating corresponding audios in several complex scenarios. However, there are instances where the model slightly miscatch the change of the scene. From our point of view, several key problems cannot be solved by the current technique, which may be ignored in the previous study: 1) the presence of an object in a video doesn't necessarily equate to it producing a sound. Some objects can produce different sounds depending on the context or the action being performed, and not all objects in a video make a sound. Models often fail to recognize these 'silent' objects in a video. Nevertheless, our model shows robust generation capabilities in open-domain tasks, generating natural and realistic audios from silent YouTube or even AI-generated videos, as shown in \cref{fig:tta_ood}. 
 
\section{Conclusion}
In this paper, we delve into the \ac{vta} task, which aims to generate semantically-related and temporally-aligned audios given silent video pieces. Leveraging the diffusion-based backbone of the \ac{vta} model, we demonstrate the effectiveness of our VTA-LDM framework and conduct various ablations in the model design and training process. We also dive into several key aspects during the \ac{vta} generation, aiming at giving the community more insights about the model design and training process. Specifically, we concentrate on the vision encoder, auxiliary embeddings, and data augmentation techniques employed in the training process, and propose several significant suggestions that have been quantitatively validated through our experiments. A combination of these approaches can lead to a more powerful \ac{vta} backbone. We foresee the evolution of more realistic and natural video-audio generation models based on the insights gained from this study.

\textbf{Limitations \& future work} The models presented in this paper are trained on a limited dataset, VGGSound, which primarily contains audios of single audio event data. Although we employ data augmentations and utilize a portion of the filtered YouTube test set, there is room for improvement. As part of our future work, we plan to build a more extensive, more diverse, and real-world-like high-quality dataset.

\textbf{Social impact} The capability to generate realistic audio from silent visual input can significantly enhance the accessibility of high-quality video-audio content, benefiting fields such as AI-generated content (AIGC) and virtual reality simulations. However, potential ethical concerns arise, including the possibility of misuse in creating deep fake videos or misleading audio content.

\bibliography{main}

\begin{thebibliography}{41}
\providecommand{\natexlab}[1]{#1}
\providecommand{\url}[1]{\texttt{#1}}
\expandafter\ifx\csname urlstyle\endcsname\relax
  \providecommand{\doi}[1]{doi: #1}\else
  \providecommand{\doi}{doi: \begingroup \urlstyle{rm}\Url}\fi

\bibitem[Arnab et~al.(2021)Arnab, Dehghani, Heigold, Sun, Lu{\v{c}}i{\'c}, and Schmid]{arnab2021vivit}
Anurag Arnab, Mostafa Dehghani, Georg Heigold, Chen Sun, Mario Lu{\v{c}}i{\'c}, and Cordelia Schmid.
\newblock Vivit: A video vision transformer.
\newblock In \emph{Proceedings of the IEEE/CVF international conference on computer vision}, pp.\  6836--6846, 2021.

\bibitem[Assran et~al.(2023)Assran, Duval, Misra, Bojanowski, Vincent, Rabbat, LeCun, and Ballas]{assran2023self}
Mahmoud Assran, Quentin Duval, Ishan Misra, Piotr Bojanowski, Pascal Vincent, Michael Rabbat, Yann LeCun, and Nicolas Ballas.
\newblock Self-supervised learning from images with a joint-embedding predictive architecture.
\newblock \emph{arXiv preprint arXiv:2301.08243}, 2023.

\bibitem[Bar-Tal et~al.(2024)Bar-Tal, Chefer, Tov, Herrmann, Paiss, Zada, Ephrat, Hur, Liu, Raj, Li, Rubinstein, Michaeli, Wang, Sun, Dekel, and Mosseri]{bartal2024lumiere}
Omer Bar-Tal, Hila Chefer, Omer Tov, Charles Herrmann, Roni Paiss, Shiran Zada, Ariel Ephrat, Junhwa Hur, Guanghui Liu, Amit Raj, Yuanzhen Li, Michael Rubinstein, Tomer Michaeli, Oliver Wang, Deqing Sun, Tali Dekel, and Inbar Mosseri.
\newblock Lumiere: A space-time diffusion model for video generation, 2024.

\bibitem[Bardes et~al.(2024)Bardes, Garrido, Ponce, Chen, Rabbat, LeCun, Assran, and Ballas]{bardes2024vjepa}
Adrien Bardes, Quentin Garrido, Jean Ponce, Xinlei Chen, Michael Rabbat, Yann LeCun, Mido Assran, and Nicolas Ballas.
\newblock V-{JEPA}: Latent video prediction for visual representation learning, 2024.
\newblock URL \url{https://openreview.net/forum?id=WFYbBOEOtv}.

\bibitem[Blattmann et~al.(2023)Blattmann, Dockhorn, Kulal, Mendelevitch, Kilian, Lorenz, Levi, English, Voleti, Letts, Jampani, and Rombach]{blattmann2023stable}
Andreas Blattmann, Tim Dockhorn, Sumith Kulal, Daniel Mendelevitch, Maciej Kilian, Dominik Lorenz, Yam Levi, Zion English, Vikram Voleti, Adam Letts, Varun Jampani, and Robin Rombach.
\newblock Stable video diffusion: Scaling latent video diffusion models to large datasets, 2023.

\bibitem[Brooks et~al.(2024)Brooks, Peebles, Holmes, DePue, Guo, Jing, Schnurr, Taylor, Luhman, Luhman, Ng, Wang, and Ramesh]{videoworldsimulators2024}
Tim Brooks, Bill Peebles, Connor Holmes, Will DePue, Yufei Guo, Li~Jing, David Schnurr, Joe Taylor, Troy Luhman, Eric Luhman, Clarence Ng, Ricky Wang, and Aditya Ramesh.
\newblock Video generation models as world simulators.
\newblock 2024.
\newblock URL \url{https://openai.com/research/video-generation-models-as-world-simulators}.

\bibitem[Chen et~al.(2020)Chen, Xie, Vedaldi, and Zisserman]{chen2020vggsound}
Honglie Chen, Weidi Xie, Andrea Vedaldi, and Andrew Zisserman.
\newblock Vggsound: A large-scale audio-visual dataset.
\newblock In \emph{ICASSP 2020-2020 IEEE International Conference on Acoustics, Speech and Signal Processing (ICASSP)}, pp.\  721--725. IEEE, 2020.

\bibitem[Deshmukh et~al.(2024)Deshmukh, Alharthi, Elizalde, Gamper, Ismail, Singh, Raj, and Wang]{deshmukh2024pam}
Soham Deshmukh, Dareen Alharthi, Benjamin Elizalde, Hannes Gamper, Mahmoud~Al Ismail, Rita Singh, Bhiksha Raj, and Huaming Wang.
\newblock Pam: Prompting audio-language models for audio quality assessment.
\newblock \emph{arXiv preprint arXiv:2402.00282}, 2024.

\bibitem[Dhariwal \& Nichol(2021)Dhariwal and Nichol]{dhariwal2021diffusion}
Prafulla Dhariwal and Alexander Nichol.
\newblock Diffusion models beat gans on image synthesis.
\newblock \emph{Advances in neural information processing systems}, 34:\penalty0 8780--8794, 2021.

\bibitem[Esser et~al.(2021)Esser, Rombach, and Ommer]{esser2021taming}
Patrick Esser, Robin Rombach, and Bjorn Ommer.
\newblock Taming transformers for high-resolution image synthesis.
\newblock In \emph{Proceedings of the IEEE/CVF conference on computer vision and pattern recognition}, pp.\  12873--12883, 2021.

\bibitem[Fedorishin et~al.(2023)Fedorishin, Mohan, Jawade, Setlur, and Govindaraju]{fedorishin2023hear}
Dennis Fedorishin, Deen~Dayal Mohan, Bhavin Jawade, Srirangaraj Setlur, and Venu Govindaraju.
\newblock Hear the flow: Optical flow-based self-supervised visual sound source localization.
\newblock In \emph{Proceedings of the IEEE/CVF Winter Conference on Applications of Computer Vision}, pp.\  2278--2287, 2023.

\bibitem[Ghosal et~al.(2023)Ghosal, Majumder, Mehrish, and Poria]{ghosal2023text}
Deepanway Ghosal, Navonil Majumder, Ambuj Mehrish, and Soujanya Poria.
\newblock Text-to-audio generation using instruction guided latent diffusion model.
\newblock In \emph{Proceedings of the 31st ACM International Conference on Multimedia}, pp.\  3590--3598, 2023.

\bibitem[Girdhar et~al.(2023)Girdhar, El-Nouby, Liu, Singh, Alwala, Joulin, and Misra]{girdhar2023imagebind}
Rohit Girdhar, Alaaeldin El-Nouby, Zhuang Liu, Mannat Singh, Kalyan~Vasudev Alwala, Armand Joulin, and Ishan Misra.
\newblock Imagebind: One embedding space to bind them all, 2023.

\bibitem[Ho \& Salimans(2022)Ho and Salimans]{ho2022classifier}
Jonathan Ho and Tim Salimans.
\newblock Classifier-free diffusion guidance.
\newblock \emph{arXiv preprint arXiv:2207.12598}, 2022.

\bibitem[Ho et~al.(2020)Ho, Jain, and Abbeel]{ho2020denoising}
Jonathan Ho, Ajay Jain, and Pieter Abbeel.
\newblock Denoising diffusion probabilistic models.
\newblock \emph{Advances in neural information processing systems}, 33:\penalty0 6840--6851, 2020.

\bibitem[Huang et~al.(2023)Huang, Ren, Huang, Yang, Ye, Zhang, Liu, Yin, Ma, and Zhao]{huang2023make2}
Jiawei Huang, Yi~Ren, Rongjie Huang, Dongchao Yang, Zhenhui Ye, Chen Zhang, Jinglin Liu, Xiang Yin, Zejun Ma, and Zhou Zhao.
\newblock Make-an-audio 2: Temporal-enhanced text-to-audio generation.
\newblock \emph{arXiv preprint arXiv:2305.18474}, 2023.

\bibitem[Iashin \& Rahtu(2021)Iashin and Rahtu]{iashin2021taming}
Vladimir Iashin and Esa Rahtu.
\newblock Taming visually guided sound generation.
\newblock \emph{arXiv preprint arXiv:2110.08791}, 2021.

\bibitem[Kong et~al.(2020)Kong, Kim, and Bae]{kong2020hifi}
Jungil Kong, Jaehyeon Kim, and Jaekyoung Bae.
\newblock Hifi-gan: Generative adversarial networks for efficient and high fidelity speech synthesis.
\newblock \emph{Advances in Neural Information Processing Systems}, 33:\penalty0 17022--17033, 2020.

\bibitem[Kreuk et~al.(2022)Kreuk, Synnaeve, Polyak, Singer, D{\'e}fossez, Copet, Parikh, Taigman, and Adi]{kreuk2022audiogen}
Felix Kreuk, Gabriel Synnaeve, Adam Polyak, Uriel Singer, Alexandre D{\'e}fossez, Jade Copet, Devi Parikh, Yaniv Taigman, and Yossi Adi.
\newblock Audiogen: Textually guided audio generation.
\newblock \emph{arXiv preprint arXiv:2209.15352}, 2022.

\bibitem[Kurmi et~al.(2021)Kurmi, Bajaj, Patro, Venkatesh, Namboodiri, and Jyothi]{kurmi2021collaborative}
Vinod~K Kurmi, Vipul Bajaj, Badri~N Patro, KS~Venkatesh, Vinay~P Namboodiri, and Preethi Jyothi.
\newblock Collaborative learning to generate audio-video jointly.
\newblock In \emph{ICASSP 2021-2021 IEEE International Conference on Acoustics, Speech and Signal Processing (ICASSP)}, pp.\  4180--4184. IEEE, 2021.

\bibitem[Liu et~al.(2023{\natexlab{a}})Liu, Chen, Yuan, Mei, Liu, Mandic, Wang, and Plumbley]{liu2023audioldm}
Haohe Liu, Zehua Chen, Yi~Yuan, Xinhao Mei, Xubo Liu, Danilo Mandic, Wenwu Wang, and Mark~D Plumbley.
\newblock Audioldm: Text-to-audio generation with latent diffusion models.
\newblock \emph{arXiv preprint arXiv:2301.12503}, 2023{\natexlab{a}}.

\bibitem[Liu et~al.(2023{\natexlab{b}})Liu, Tian, Yuan, Liu, Mei, Kong, Wang, Wang, Wang, and Plumbley]{liu2023audioldm2}
Haohe Liu, Qiao Tian, Yi~Yuan, Xubo Liu, Xinhao Mei, Qiuqiang Kong, Yuping Wang, Wenwu Wang, Yuxuan Wang, and Mark~D Plumbley.
\newblock Audioldm 2: Learning holistic audio generation with self-supervised pretraining.
\newblock \emph{arXiv preprint arXiv:2308.05734}, 2023{\natexlab{b}}.

\bibitem[Luo et~al.(2021)Luo, Ji, Zhong, Chen, Lei, Duan, and Li]{luo2021clip4clip}
Huaishao Luo, Lei Ji, Ming Zhong, Yang Chen, Wen Lei, Nan Duan, and Tianrui Li.
\newblock Clip4clip: An empirical study of clip for end to end video clip retrieval, 2021.

\bibitem[Luo et~al.(2024)Luo, Yan, Hu, and Zhao]{luo2024diff}
Simian Luo, Chuanhao Yan, Chenxu Hu, and Hang Zhao.
\newblock Diff-foley: Synchronized video-to-audio synthesis with latent diffusion models.
\newblock \emph{Advances in Neural Information Processing Systems}, 36, 2024.

\bibitem[Mei et~al.(2023)Mei, Meng, Liu, Kong, Ko, Zhao, Plumbley, Zou, and Wang]{mei2023wavcaps}
Xinhao Mei, Chutong Meng, Haohe Liu, Qiuqiang Kong, Tom Ko, Chengqi Zhao, Mark~D Plumbley, Yuexian Zou, and Wenwu Wang.
\newblock Wavcaps: A chatgpt-assisted weakly-labelled audio captioning dataset for audio-language multimodal research.
\newblock \emph{arXiv preprint arXiv:2303.17395}, 2023.

\bibitem[Mo et~al.(2024)Mo, Shi, and Tian]{mo2024text}
Shentong Mo, Jing Shi, and Yapeng Tian.
\newblock Text-to-audio generation synchronized with videos.
\newblock \emph{arXiv preprint arXiv:2403.07938}, 2024.

\bibitem[Nichol et~al.(2021)Nichol, Dhariwal, Ramesh, Shyam, Mishkin, McGrew, Sutskever, and Chen]{nichol2021glide}
Alex Nichol, Prafulla Dhariwal, Aditya Ramesh, Pranav Shyam, Pamela Mishkin, Bob McGrew, Ilya Sutskever, and Mark Chen.
\newblock Glide: Towards photorealistic image generation and editing with text-guided diffusion models.
\newblock \emph{arXiv preprint arXiv:2112.10741}, 2021.

\bibitem[Radford et~al.(2021)Radford, Kim, Hallacy, Ramesh, Goh, Agarwal, Sastry, Askell, Mishkin, Clark, et~al.]{radford2021learning}
Alec Radford, Jong~Wook Kim, Chris Hallacy, Aditya Ramesh, Gabriel Goh, Sandhini Agarwal, Girish Sastry, Amanda Askell, Pamela Mishkin, Jack Clark, et~al.
\newblock Learning transferable visual models from natural language supervision.
\newblock In \emph{International conference on machine learning}, pp.\  8748--8763. PMLR, 2021.

\bibitem[Rombach et~al.(2022)Rombach, Blattmann, Lorenz, Esser, and Ommer]{rombach2022high}
Robin Rombach, Andreas Blattmann, Dominik Lorenz, Patrick Esser, and Bj{\"o}rn Ommer.
\newblock High-resolution image synthesis with latent diffusion models.
\newblock In \emph{Proceedings of the IEEE/CVF conference on computer vision and pattern recognition}, pp.\  10684--10695, 2022.

\bibitem[Ruan et~al.(2023)Ruan, Ma, Yang, He, Liu, Fu, Yuan, Jin, and Guo]{ruan2023mm}
Ludan Ruan, Yiyang Ma, Huan Yang, Huiguo He, Bei Liu, Jianlong Fu, Nicholas~Jing Yuan, Qin Jin, and Baining Guo.
\newblock Mm-diffusion: Learning multi-modal diffusion models for joint audio and video generation.
\newblock In \emph{Proceedings of the IEEE/CVF Conference on Computer Vision and Pattern Recognition}, pp.\  10219--10228, 2023.

\bibitem[Sheffer \& Adi(2023)Sheffer and Adi]{sheffer2023hear}
Roy Sheffer and Yossi Adi.
\newblock I hear your true colors: Image guided audio generation.
\newblock In \emph{ICASSP 2023-2023 IEEE International Conference on Acoustics, Speech and Signal Processing (ICASSP)}, pp.\  1--5. IEEE, 2023.

\bibitem[Singer et~al.(2022)Singer, Polyak, Hayes, Yin, An, Zhang, Hu, Yang, Ashual, Gafni, et~al.]{singer2022make}
Uriel Singer, Adam Polyak, Thomas Hayes, Xi~Yin, Jie An, Songyang Zhang, Qiyuan Hu, Harry Yang, Oron Ashual, Oran Gafni, et~al.
\newblock Make-a-video: Text-to-video generation without text-video data.
\newblock \emph{arXiv preprint arXiv:2209.14792}, 2022.

\bibitem[Vaswani et~al.(2023)Vaswani, Shazeer, Parmar, Uszkoreit, Jones, Gomez, Kaiser, and Polosukhin]{vaswani2023attention}
Ashish Vaswani, Noam Shazeer, Niki Parmar, Jakob Uszkoreit, Llion Jones, Aidan~N. Gomez, Lukasz Kaiser, and Illia Polosukhin.
\newblock Attention is all you need, 2023.

\bibitem[Wang et~al.(2024)Wang, Ma, Pascual, Cartwright, and Cai]{v2a-mapper}
Heng Wang, Jianbo Ma, Santiago Pascual, Richard Cartwright, and Weidong Cai.
\newblock V2a-mapper: A lightweight solution for vision-to-audio generation by connecting foundation models.
\newblock In \emph{Proceedings of the AAAI Conference on Artificial Intelligence}, 2024.

\bibitem[Wu* et~al.(2023)Wu*, Chen*, Zhang*, Hui*, Berg-Kirkpatrick, and Dubnov]{laionclap2023}
Yusong Wu*, Ke~Chen*, Tianyu Zhang*, Yuchen Hui*, Taylor Berg-Kirkpatrick, and Shlomo Dubnov.
\newblock Large-scale contrastive language-audio pretraining with feature fusion and keyword-to-caption augmentation.
\newblock In \emph{IEEE International Conference on Acoustics, Speech and Signal Processing, ICASSP}, 2023.

\bibitem[Xing et~al.(2024)Xing, He, Tian, Wang, and Chen]{xing2024seeing}
Yazhou Xing, Yingqing He, Zeyue Tian, Xintao Wang, and Qifeng Chen.
\newblock Seeing and hearing: Open-domain visual-audio generation with diffusion latent aligners.
\newblock \emph{arXiv preprint arXiv:2402.17723}, 2024.

\bibitem[Xu et~al.(2024)Xu, Li, Su, Liang, Yu, et~al.]{xu2024prompt}
Manjie Xu, Chenxing Li, Dan Su, Wei Liang, Dong Yu, et~al.
\newblock Prompt-guided precise audio editing with diffusion models.
\newblock \emph{arXiv preprint arXiv:2406.04350}, 2024.

\bibitem[Yariv et~al.(2024)Yariv, Gat, Benaim, Wolf, Schwartz, and Adi]{yariv2024diverse}
Guy Yariv, Itai Gat, Sagie Benaim, Lior Wolf, Idan Schwartz, and Yossi Adi.
\newblock Diverse and aligned audio-to-video generation via text-to-video model adaptation.
\newblock In \emph{Proceedings of the AAAI Conference on Artificial Intelligence}, volume~38, pp.\  6639--6647, 2024.

\bibitem[Zeng et~al.(2023)Zeng, Wei, Zheng, Zou, Wei, Zhang, and Li]{zeng2023make}
Yan Zeng, Guoqiang Wei, Jiani Zheng, Jiaxin Zou, Yang Wei, Yuchen Zhang, and Hang Li.
\newblock Make pixels dance: High-dynamic video generation, 2023.

\bibitem[Zhang et~al.(2024)Zhang, Gu, Zeng, Xing, Wang, Wu, and Chen]{zhang2024foleycrafter}
Yiming Zhang, Yicheng Gu, Yanhong Zeng, Zhening Xing, Yuancheng Wang, Zhizheng Wu, and Kai Chen.
\newblock Foleycrafter: Bring silent videos to life with lifelike and synchronized sounds.
\newblock 2024.

\bibitem[Zhu et~al.(2023)Zhu, Lin, Ning, Yan, Cui, HongFa, Pang, Jiang, Zhang, Li, Zhang, Li, Liu, and Yuan]{zhu2023languagebind}
Bin Zhu, Bin Lin, Munan Ning, Yang Yan, Jiaxi Cui, Wang HongFa, Yatian Pang, Wenhao Jiang, Junwu Zhang, Zongwei Li, Cai~Wan Zhang, Zhifeng Li, Wei Liu, and Li~Yuan.
\newblock Languagebind: Extending video-language pretraining to n-modality by language-based semantic alignment, 2023.

\end{thebibliography}
\bibliographystyle{iclr2025_conference}

\newpage
\appendix

\section{Baselines}
\label{supp_baselines}
We compare with these open-sourced \ac{vta} baselines:

\begin{enumerate}
    \item IM2WAV~\citep{sheffer2023hear} is an open-domain audio
    generation system which is based on the image or the image sequence. The model uses a language model to generate low-level audio representation. We use the pre-trained checkpoint with the default parameters as the baseline.  
    \item Diff-Foley~\citep{luo2024diff} uses a \ac{ldm} based on the features extracted by CAVP, an encoder that is contrastively pre-trained to learn temporally and semantically aligned audio-visual features. We use the pre-trained Diff-Foley as our baseline.
    \item FoleyCrafter \citep{zhang2024foleycrafter} imports semantic and temporal blocks for precise audio-video synchronization, and supports the use of text descriptions to facilitate controllable and diverse \ac{tta} generation.
    \item Seeing\&Hearing~\citep{xing2024seeing} is built on a multimodality latent aligner with the pre-trained ImageBind model. It also uses a \ac{ldm} as the generation framework.
    \item T2AV~\citep{mo2024text} leverages visual-aligned text embeddings as its conditional foundation in diffusion-based audio generation. Note that, compared with other baselines, T2AV generation is more based on the textual description. The model leverages a pre-trained video-audio \ac{clap}(VA-CLAP) as the vision encoder.

\end{enumerate}

\section{Subjection Evaluation}
\label{sup_subjective}
The subjective evaluation aims to gauge the performance of our models from a human perspective, providing insights into the perceived quality and alignment of the generated audio and original video content. Participants undergo a training session where they are introduced to the evaluation objectives, the rating scale, and example demonstrations that illustrate different quality levels. Additionally, they practice rating sample videos to familiarize themselves with the criteria and receive feedback to ensure consistency.

\paragraph{Overall Quality} This metric evaluates the general appeal and coherence of the combined video, considering both audio and visual components. It contains how well the audio and video elements fit together without any jarring or inconsistent moments.

\paragraph{Audio Quality} Focusing solely on the generated audio, this metric assesses factors such as clarity, fidelity, and naturalness, like how clear and understandable the audio is, free from distortion or muddiness, and how closely the generated audio resembles human speech or natural sounds in terms of intonation, rhythm, and expression.

\paragraph{Video-Audio Semantic Alignment} This metric measures how well the audio semantically matches the visual content, ensuring that the sounds correspond appropriately to the actions and scenes depicted in the video, like how relevant the audio is to the visual context, with sounds matching the actions and scenes on screen.

\paragraph{Video-Audio Temporal Alignment} The evaluation centers on the synchronization between the audio and video streams, determining how accurately the timing of audio events aligns with visual events, like the precision with which sounds occur in tandem with corresponding visual actions (e.g., a door slamming in sync with the sound of the slam).

\section{Experiments on Vision Encoders}
\label{supp_encoder}
Based on the proposed generation framework, we start by trying ablation results on different vision encoders:
\begin{enumerate}
    \item Clip4Clip~\citep{luo2021clip4clip} foucuses on video-text retrieval based on the pretrained \ac{clip}~\citep{radford2021learning} model. Experiments have shown that CLIP can serve as the backbone to extract knowledge from frame-level video input. We leverage a similar linear projection to map the frame-level video features extracted by CLIP to get the generation condition. 
    \item Imagebind~\citep{girdhar2023imagebind} is pre-trained on different modalities to learn joint embeddings, including images, text, and audio. All the modalities are bound with the image-paired data. We leverage Imagebind to get the image embeddings of the extracted frames extracted from the input video. 
    \item LanguageBind~\citep{zhu2023languagebind} is trained similarly with Imagebind but takes the language as the bind across different modalities. As the language modality is well-explored and proved influential as conditions in the previous \ac{tta} works, we would like to see whether vision embeddings extracted from LanguageBind can serve as a better condition.
    \item V-JEPA~\citep{bardes2024vjepa} is considered as an extension of the I-JEPA~\citep{assran2023self} to video based on self-supervised learning. Improving from I-JEPA which learns semantic image features, V-JEPA deals with self-supervised learning of video representations appropriate for video understanding in a spatio-temporal way.
    \item ViViT~\citep{arnab2021vivit} a pure-transformer based models. Although the model was originally designed for video classification, we would like to check whether the spatio-temporal tokens learned from the input video can help guide the generation of audio. We use the pre-trained ViViT trained on the kinetics400 dataset as the encoder.
    \item CAVP~\citep{luo2024diff} is trained with two different objects, semantic contrast loss and temporal contrast loss, to improve audio-video features' semantic and temporal alignment. CAVP was first leveraged in the DIFF-FOLEY model to synthesize synchronized video-conditioned content. We use the pre-trained CAVP (also pre-trained on the VGGSound dataset) as our vision encoder. 
\end{enumerate}

\section{Experiments on Auxiliary Embeddings}
\label{supp_aux}
We are also interested in exploring whether auxiliary embeddings, beyond visual features, could enhance the generation process. Numerous studies have demonstrated that additional information can improve generation results in various ways, whether broadly or specifically. We aim to investigate this phenomenon through several auxiliary embeddings:
\begin{enumerate}
    \item \textbf{Text Information}: Additional textual labels can provide valuable context and extra semantic details that may not be immediately discernible from visual features. Moreover, text information can assist in filtering out extraneous information in the video. For instance, the presence of a gun in a video does not necessarily imply that the corresponding audio will be generated. In our experiment, we utilize CLIP to obtain text embeddings and concatenate these with the video embeddings to serve as the condition.
    \item \textbf{Position Embedding}: Indeed, Position Embedding is critical as it imparts a sense of temporal order or sequence, which is essential in audio generation, particularly when the vision encoders are primarily focused on the semantics of the audio. Intuitively, Position Embedding assists the model in understanding the event sequence, thereby enabling it to generate coherent audio. We utilize sinusoidal positional embedding, akin to the method used in the Transformer model~\citep{vaswani2023attention}, as it facilitates effortless attention to relative positions.
    \item \textbf{Optical Flow}: Optical Flow offers valuable insights into the motion and dynamics present in a video sequence. Previous study~\citep{fedorishin2023hear} has utilized this information to assist in localizing sound sources within videos. In our approach, we employ optical flow video embeddings as an additional condition for generation. Similar to the previous embeddings, these embeddings are concatenated with the original video embeddings, enriching the input representation and potentially enhancing the audio generation process.
    
\end{enumerate}

\section{Experiments on Data Augmentation}
\label{supp_dataaug}
The data quality of the training set is undeniably crucial for a model's performance, particularly when training large generative models. We explore data augmentation from three different perspectives:
\begin{enumerate}
    \item \textbf{Data Clean}: Data clean ensures that the input data is accurate, consistent, and free from errors or anomalies. We use a CLAP~\citep{laionclap2023} model to help select audio-video pairs with similar semantics based on extra textual labels (with $score>0.3$). We also use the AV-Align score to filter unmatched video-audio pairs (with $score>0.2$). For the VGGSound dataset, we filter out about 100k high-quality video-audio pairs.
    
    \item \textbf{Concat Augmnent} The original VGGSound dataset primarily contains audio clips with single audio events. While this makes the dataset a clean, simple test set for audio event generation, it does not evaluate the model's ability to handle temporal information in complex videos. To simulate complex generation tasks with various audio events during training, we randomly concatenate two videos with different audio events. We also propose a test set to assess our models.

    \item \textbf{Pretrain} We also try to leverage the power of pretraining. We pretrain our model on two different data corpus seperately: a large video-audio corpus consisting of 10k hours of content from YouTube, and a large audio corpus mainly from WavCaps~\citep{mei2023wavcaps} and Youtube, amounting to approximately 150k hours of audio with paired captions. The YouTube corpus is processed in the same manner as the VGGSound video data. We filter out talking and music content, using only audio event cases for training. For video-audio pretraining, we leverage video-to-audio supervised training. For audio pretraining, we perform audio self-supervised training. Subsequently, we perform full parameter finetuning of our model on the original training set.
    
\end{enumerate}

\section{More Demos}
\subsection{Sailency Map}
\begin{figure*}[h!]
    \centering
    \small
    \includegraphics[width=\linewidth]{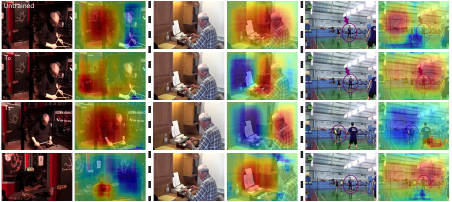}
    \caption{\textbf{More demos of the sailency map.} We show that the model focuses on different components in different frames of the given video after training, although the vision encoder parameters are completely freezed during the training time.}
    \label{fig:supp_sailency}
\end{figure*}

\subsection{Audio Generation}
\begin{figure*}[h!]
    \centering
    \small
    \includegraphics[width=\linewidth]{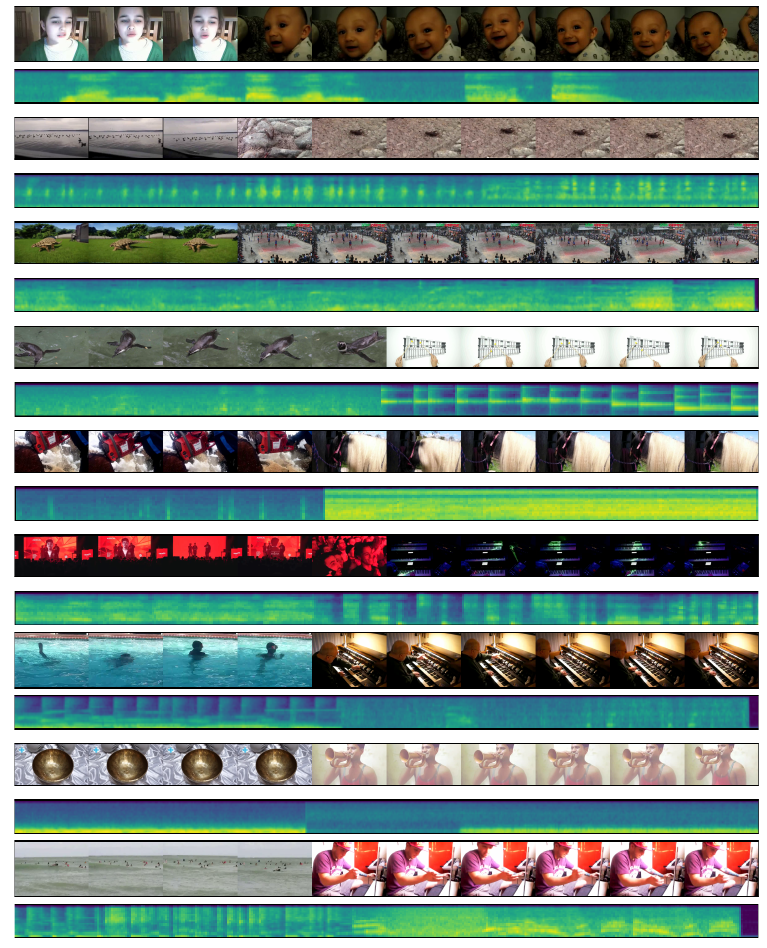}
    \caption{\textbf{More demos of the \ac{vta} generation on open-domain videos.} Refer to the supp materials for more video demos and the inference code.}
    \label{fig:supp_new}
\end{figure*}

\end{document}